\documentclass[reprint,aps,prb,twocolumn,longbibliography,superscriptaddress]{revtex4-1}
\pdfoutput=1
\usepackage{graphicx}
\usepackage{subfigure}
\usepackage{epstopdf}
\usepackage{lineno}
\usepackage{bm}
\usepackage{verbatim}
\usepackage{dcolumn}
\usepackage{amssymb}
\usepackage{amsmath}
\usepackage{gensymb}
\usepackage{array,multirow}
\usepackage{dcolumn}
\usepackage[normalem]{ulem}
\usepackage{cancel}
\usepackage{afterpage}
\usepackage{xfrac}
\usepackage{xcolor}
\usepackage[sort&compress]{natbib}
\usepackage[colorlinks=true,allcolors=blue]{hyperref}
\begin{document}
\title{Coexistence of Weyl Physics and Planar Defects in Semimetals TaP and TaAs}

\author{T. Besara}
    \email{besara@magnet.fsu.edu}
    \affiliation{National High Magnetic Field Laboratory, Florida State University, Tallahassee, FL 32310, USA}
\author{D. A. Rhodes}
    \affiliation{National High Magnetic Field Laboratory, Florida State University, Tallahassee, FL 32310, USA}
\author{K.-W. Chen}
    \affiliation{National High Magnetic Field Laboratory, Florida State University, Tallahassee, FL 32310, USA}
\author{S. Das}
    \affiliation{National High Magnetic Field Laboratory, Florida State University, Tallahassee, FL 32310, USA}
    \affiliation{Department of Physics, Florida State University, Tallahassee, FL 32306, USA}
\author{Q. R. Zhang}
    \affiliation{National High Magnetic Field Laboratory, Florida State University, Tallahassee, FL 32310, USA}
\author{J. Sun}
    \affiliation{Department of Physics and Astronomy, University of Missouri-Columbia, Columbia, MO 65211, USA}
\author{B. Zeng}
    \affiliation{National High Magnetic Field Laboratory, Florida State University, Tallahassee, FL 32310, USA}
\author{Y. Xin}
    \affiliation{National High Magnetic Field Laboratory, Florida State University, Tallahassee, FL 32310, USA}
\author{L. Balicas}
    \affiliation{National High Magnetic Field Laboratory, Florida State University, Tallahassee, FL 32310, USA}
\author{R. E. Baumbach}
    \affiliation{National High Magnetic Field Laboratory, Florida State University, Tallahassee, FL 32310, USA}
\author{E. Manousakis}
    \affiliation{National High Magnetic Field Laboratory, Florida State University, Tallahassee, FL 32310, USA}
    \affiliation{Department of Physics, Florida State University, Tallahassee, FL 32306, USA}
\author{D. J. Singh}
    \affiliation{Department of Physics and Astronomy, University of Missouri-Columbia, Columbia, MO 65211, USA}
\author{T. Siegrist}
    \affiliation{National High Magnetic Field Laboratory, Florida State University, Tallahassee, FL 32310, USA}
    \affiliation{Department of Chemical and Biomedical Engineering, FAMU-FSU College of Engineering, Florida State University, Tallahassee, FL 32310, USA}

\date{\today}

\begin{abstract}
We report a structural study of the Weyl semimetals TaAs and TaP, utilizing diffraction and imaging techniques, where we show that they contain a high density of defects, leading to non-stoichiometric single crystals of both semimetals. Despite the observed defects and non-stoichiometry on samples grown using techniques already reported in the literature, de Haas-van Alphen measurements on TaP reveal quantum oscillations and a high carrier mobility, an indication that the crystals are of quality comparable to those reported elsewhere. Electronic structure calculations on TaAs reveal that the position of the Weyl points relative to the Fermi level shift with the introduction of vacancies and stacking faults. In the case of vacancies the Fermi surface becomes considerably altered, while the effect of stacking faults on the electronic structure is to allow the Weyl pockets to remain close to the Fermi surface. The observation of quantum oscillations in a non-stoichiometric crystal and the persistence of Weyl fermion pockets near the Fermi surface in a crystal with stacking faults point to the robustness of these quantum phenomena in these materials.
\end{abstract}

\maketitle

\section{Introduction}
Weyl fermions, massless fermions predicted by Hermann Weyl in 1929\cite{Weyl_ZP_1929} as solutions to the Dirac equation, have not yet been observed as fundamental particles in high energy physics. In 2011, however, it was predicted that Weyl fermions can be realized in condensed matter physics as electronic quasi-particles in the family of pyrochlore iridates\cite{Wan_PRB_2011} and the ferromagnetic spinel compound HgCr$_2$Se$_4$.\cite{Xu_PRL_2011}

Following recent theoretical predictions of Weyl fermions in the simple semimetal TaAs and its isostructural compounds TaP, NbAs, and NbP,\cite{Weng_PRX_2015, Huang_NatComm_2015} Weyl fermions were discovered experimentally shortly thereafter in TaAs\cite{Xu_Science_2015, Lv_PRX_2015} and was quickly confirmed by additional studies,\cite{Lv_NatPhys_2015, Yang_NatPhys_2015} along with the discovery of Weyl fermions in the isostructural NbAs\cite{Xu_NatPhys_2015} and TaP.\cite{Xu_SciAdv_2015} Several studies on these semimetals have emerged: detailed investigations of the Fermi surface topology,\cite{Lee_PRB_2015, Sun_PRB_2015, Lv_PRL_2015} the observation of large magnetoresistance and high carrier mobility,\cite{Zhang_PRB_2015, Huang_PRX_2015, Shekhar_NatPhys_2015, Ghimire_JPCM_2015, Zhang_ARXIV_2015a, Arnold_ARXIV_2015, Wang_PRB_2016} the report of a quantum phase transition in TaP,\cite{Zhang_ARXIV_2015} a Raman study of the lattice dynamics identifying all optical phonon modes in TaAs,\cite{Liu_PRB_2015} a magnetization study of TaAs,\cite{Liu_JMMM_2016} and pressure studies of NbAs.\cite{Zhang_CPL_2015, Luo_JPCM_2016}\\
\\
Until now, all four transition metal pnictide semimetals -- TaP, TaAs, NbP, and NbAs -- have been studied with an \emph{assumed} nominal 1:1 stoichiometric ratio between the transition metal and the pnictide. However, it is well known that the thermodynamic and transport properties of a material depend on the actual stoichiometry: e.g., the magnetoresistance and the Fermi surface topology might be modified by disorder. In fact, a recent study observed quantum interference patterns arising from quasi-particle scattering near point defects on the surface of a single crystalline NbP,\cite{Zheng_ACSN_2016} followed by a theoretical investigation of surface state quasi-particle interference patterns in TaAs and NbP.\cite{Chang_PRL_2016} It is therefore pertinent that a structural study be carried out on these materials to determine what, if any, defects are present.

The four semimetals have been extensively studied prior to the recent surge of interest. In fact, a non-stoichiometric composition for one of these compounds was already reported in 1954:\cite{Schonberg_ACS_1954} niobium phosphide was found as NbP$_{0.95}$, and an assumption was made that TaP would have a similar composition: TaP$_{0.95}$. The reported symmetry, the centrosymmetric group $I4_{1}/amd$, was later corrected to the non-centrosymmetric space group $I4_{1}md$ (\#~109) for all isostructural semimetals.\cite{Boller_ActaCryst_1963, Furuseth_ActaCryst_1964} The phase relations were further explored in a number of reports.\cite{Furuseth_Nature_1964, Saini_CanJChem_1964, Furuseth_ActaChemScand_1964, Furuseth_ActaChemScand_1965, Rundqvist_Nature_1966, Murray_JLess_1976} Stacking disorder was reported by Willerstr\"{o}m in all four semimetals,\cite{Willerstrom_JLess_1984} where the disorder originates from a formation of a metastable WC-type (hexagonal, $P\bar{6}m2$) during the early stages of the synthesis reaction that would partially transform into the stable NbAs-type structure, resulting in a structure containing variable fractions of NbAs- and WC-type moieties. Depending on the temperature at which the powder samples were removed from the furnace, the nominal composition changed. Xu \emph{et al.}\cite{Xu_InorgChem_1996} performed an extensive study on the crystal structure, electrical transport, and magnetic properties of single crystalline NbP. However, it was reported as stoichiometric with no defects of any kind. Saparov \emph{et al.}\cite{Saparov_SST_2012} reported on extensive structure, thermodynamic and transport properties on a series of transition metal arsenides, including TaAs and NbAs, but the samples were not single crystalline.\\
\\
Here, we report a structural study on the tantalum pnictide semimetals TaAs and TaP utilizing single crystal x-ray diffraction (XRD), energy dispersive spectroscopy (EDS), and transmission electron microscopy/scanning transmission electron microscopy (TEM/STEM).
\begin{figure}[t]
    \begin{center}
        \includegraphics[width=.75\columnwidth]{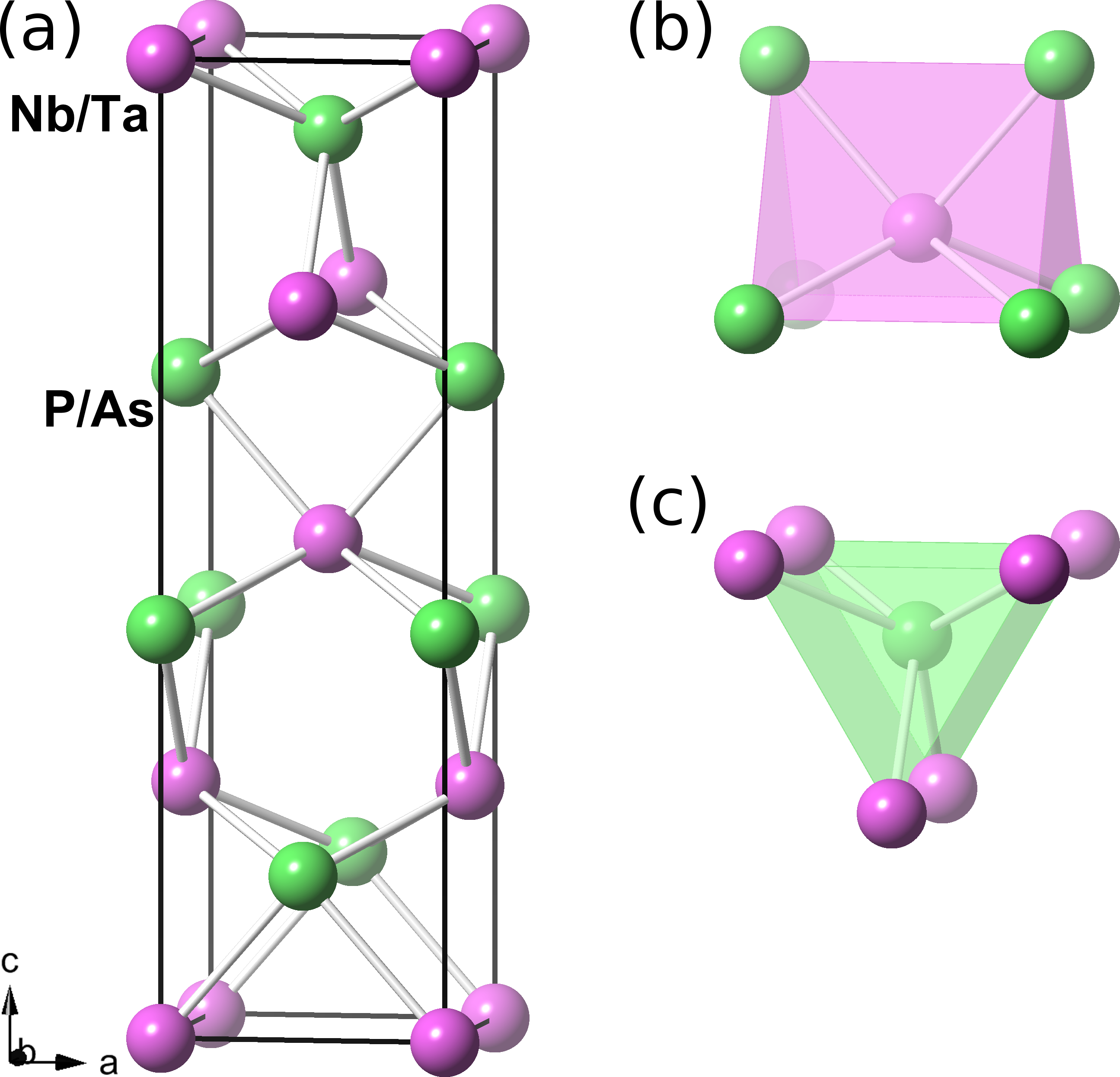}
    \end{center}
        \caption{(Color online) Crystal structure of \textit{TMPn} (\textit{TM}=Ta, Nb, and \textit{Pn}=As, P), which crystallizes in the non-centrosymmetric space group $I4_{1}md$, depicting (a) the unit cell, and the local environment of (b) the transition metal, and (c) the pnictide.}
        \label{fig:struct}
\end{figure}

XRD is essentially a measure of electron density and provides detailed information on the stoichiometry of a single crystal. However, it is not as sensitive to defects as other methods, as it assumes that all intensity is located in Bragg peaks. Therefore, point defects show up as reduced electron density, and stacking faults can result in apparent twinning or in the formation of anti-sites or anti-domains. Both of these will affect the measured electron density. EDS cannot give any information on possible defects, but it gives a value of the composition of a single crystal and therefore the overall stoichiometry, although it often has a large margin of error. In order to be able to detect defects, TEM and STEM can give a detailed picture of the atomic arrangement in a crystal, and, therefore, this technique can clearly identify any possible point and planar defects.\\
\\
Utilizing these three experimental methods, we show that these semimetals -- synthesized using techniques recently reported in the literature -- are, in fact, non-stoichiometric, and display substantial defect densities. The defects manifest themselves as site vacancies, anti-site disorder and anti-domains due to stacking faults. In TaP, we observe a phosphorous deficiency in both XRD and EDS accompanied by stacking faults, anti-site disorder and vacancies. In contrast, for TaAs, we observe a tantalum deficiency in XRD but a range of deficiencies in EDS, accompanied by a high density of stacking faults only.

In addition, we present electronic structure calculations for TaAs where we show the effect of vacancies and stacking faults on the band structure and Fermi surface. Furthermore, we present de Haas-van Alphen measurements (magnetic torque) on TaP to demonstrate that the specimens studied here are of similar quality to those discussed elsewhere.

\section{Results \& Discussion}
The (Nb,Ta)(P,As) materials crystallize in the space group $I4_{1}md$ with the structure built up by a three dimensional network of trigonal prisms of \textit{TMPn}$_6$ and \textit{PnTM}$_6$ (\textit{TM}=Nb, Ta and \textit{Pn}=P, As), as can be seen in Fig.~\ref{fig:struct}. The results of the structural refinements from the single crystal x-ray diffraction of the two tantalum semimetals are summarized in Table~\ref{tbl:xrd}. Both semimetals refine with a Flack parameter\cite{Flack_ActaCrystA_1983} of approximately 0.5, indicating that these materials are \emph{racemic compounds}. What follows are results for each semimetal.\\
\\
\begin{table}[b]
    \begin{center}
        \caption{Single crystal x-ray diffraction refinement parameters of the Ta\textit{Pn} semimetals, collected at ambient temperature. The semimetals crystallize in $I4_{1}md$ with $Z=4$. The atomic parameters and anisotropic displacement parameters (in $\times10^{4}$ \textrm{\AA}$^2$) are at the bottom. The atomic $x$- and $y$-coordinates are the same for all atoms. The Ta $z$-coordinate at 4a ($0,0,z$) is fixed at $z=0$, whereas the \textit{Pn} $z$-coordinate is refined.}
        \begin{tabular}{l c c c|c c|c c}
            \hline
            \multicolumn{4}{l|}{} & \multicolumn{2}{c|}{\textbf{TaP}} & \multicolumn{2}{c}{\textbf{TaAs}} \\
            \hline
            \multicolumn{4}{l|}{\textbf{Composition}} & \multicolumn{2}{l|}{\textbf{TaP}$\mathbf{_{0.83(3)}}$} & \multicolumn{2}{l}{\textbf{Ta}$\mathbf{_{0.92(2)}}$\textbf{As}} \\
            \multicolumn{4}{l|}{Formula weight (g/mol)} & \multicolumn{2}{l|}{206.76} & \multicolumn{2}{l}{241.21} \\
            \multicolumn{4}{l|}{$a$ (\textrm{\AA})} & \multicolumn{2}{l|}{3.31641(5)} & \multicolumn{2}{l}{3.43646(7)} \\
            \multicolumn{4}{l|}{$c$ (\textrm{\AA})} & \multicolumn{2}{l|}{11.3353(2)} & \multicolumn{2}{l}{11.6417(3)} \\
            \multicolumn{4}{l|}{$c/a$} & \multicolumn{2}{l|}{3.4179(1)} & \multicolumn{2}{l}{3.3877(1)} \\
            \multicolumn{4}{l|}{Volume (\textrm{\AA}$^{3}$)} & \multicolumn{2}{l|}{124.672(2)} & \multicolumn{2}{l}{137.480(4)} \\
            \multicolumn{4}{l|}{$\rho_{\textrm{calc}}$ (g/cm$^{3}$)} & \multicolumn{2}{l|}{11.015} & \multicolumn{2}{l}{11.653} \\
            \multicolumn{4}{l|}{Data collection range} & \multicolumn{2}{l|}{$6.4\degree\leq\theta\leq66.4\degree$} & \multicolumn{2}{l}{$6.2\degree\leq\theta\leq66.2\degree$} \\
            \multicolumn{4}{l|}{Reflections collected} & \multicolumn{2}{l|}{3409} & \multicolumn{2}{l}{2562} \\
            \multicolumn{4}{l|}{Independent reflections} & \multicolumn{2}{l|}{657} & \multicolumn{2}{l}{714} \\
            \multicolumn{4}{l|}{Parameters refined} & \multicolumn{2}{l|}{11} & \multicolumn{2}{l}{11} \\
            \multicolumn{4}{l|}{$R_{1}$, $wR_{2}$} & \multicolumn{2}{l|}{0.0543, 0.0971} & \multicolumn{2}{l}{0.0527, 0.0984} \\
            \multicolumn{4}{l|}{Goodness-of-fit on $F^{2}$} & \multicolumn{2}{l|}{0.9999} & \multicolumn{2}{l}{1.0000} \\
            \hline
            \multicolumn{4}{l|}{} & \multicolumn{2}{l|}{} & \multicolumn{2}{l}{} \\
            \hline
            Atom & Site & $x$ & $y$ & $z$ & $U_{\textrm{eq}}$ & $z$ & $U_{\textrm{eq}}$ \\
            \hline
            Ta               & 4a & 0 & 0 & 0         & 30(4) & 0         & 15(3) \\
            \emph{Pn}=As, P  & 4a & 0 & 0 & 0.4173(3) & 16(2) & 0.4173(2) & 25(6) \\
            \hline
        \end{tabular}
        \label{tbl:xrd}
    \end{center}
\end{table}
\begin{figure*}[t]
    \begin{center}
        \includegraphics[width=1.0\textwidth]{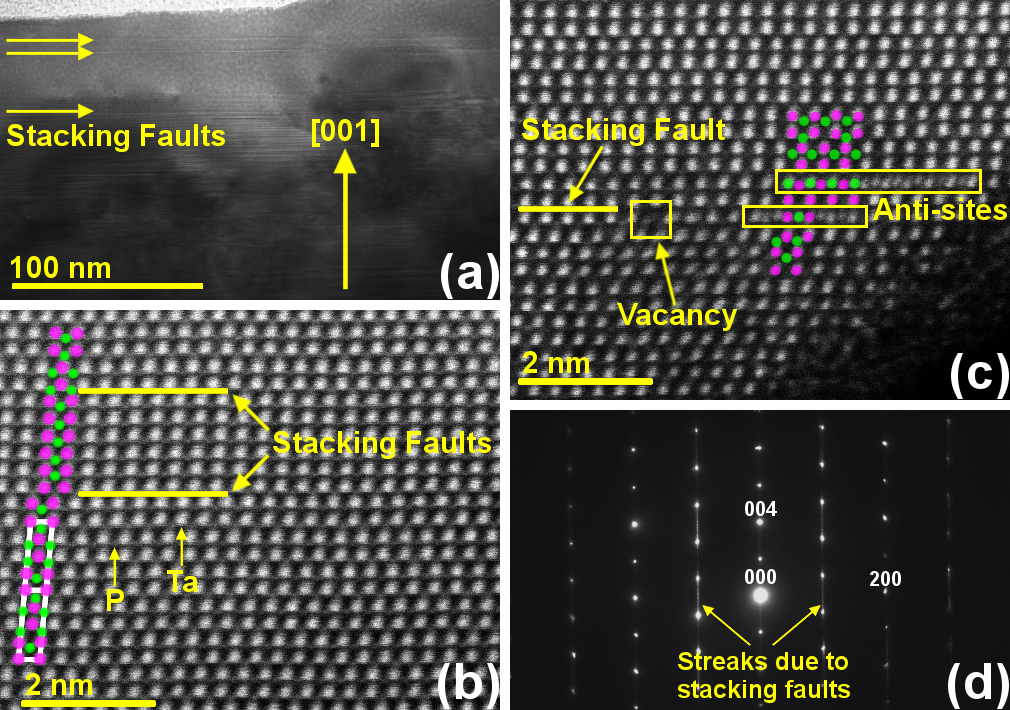}
    \end{center}
        \caption{(Color online) TEM/STEM images of TaP (the tilt in the images is due to drift). (a) TEM bright field image view of a TaP crystal with arrows highlighting the stacking faults. The faults are stacked along the $c$-axis. (b) Atomic resolution HAADF-STEM $Z$-contrast image viewed along $[010]$, with a region of stacking faults. As indicated by arrows, the bright spheres are Ta atoms while the smaller, less intense spheres are P atoms. Atoms along a strip in the $c$-direction have been highlighted with different colors to display the atomic models of the known $I4_{1}md$ structure and the stacking fault region. For clarity, the model atoms of Ta and P are of the same size. (c) An HAADF-STEM image of a different region, highlighting anti-sites, a site vacancy, and a stacking fault. A selection of atoms has been colored to display the atomic model across the anti-sites. (d) Diffraction pattern of the $[010]$ direction showing streaks arising due to the stacking faults.}
        \label{fig:taptem}
\end{figure*}
\textbf{TaP.} The single crystal refinement of the TaP crystal structure displayed larger than expected anisotropic displacement parameters (ADPs) for phosphorus when compared to tantalum ADPs, indicating P-site deficiency. Refining the site occupancy factor (SOF) of the P-site indeed yielded a significant drop in occupancy to 0.83, and a commensurate reduction of the anisotropic displacement parameters to match those of Ta. This scenario was repeated for a different crystal and gave identical results within errors. Elemental analysis via EDS on single crystals reveals a stoichiometric range of TaP$_{0.82-0.84}$, in excellent agreement with the XRD results.
\begin{figure}[t]
    \begin{center}
        \includegraphics[width=1.0\columnwidth]{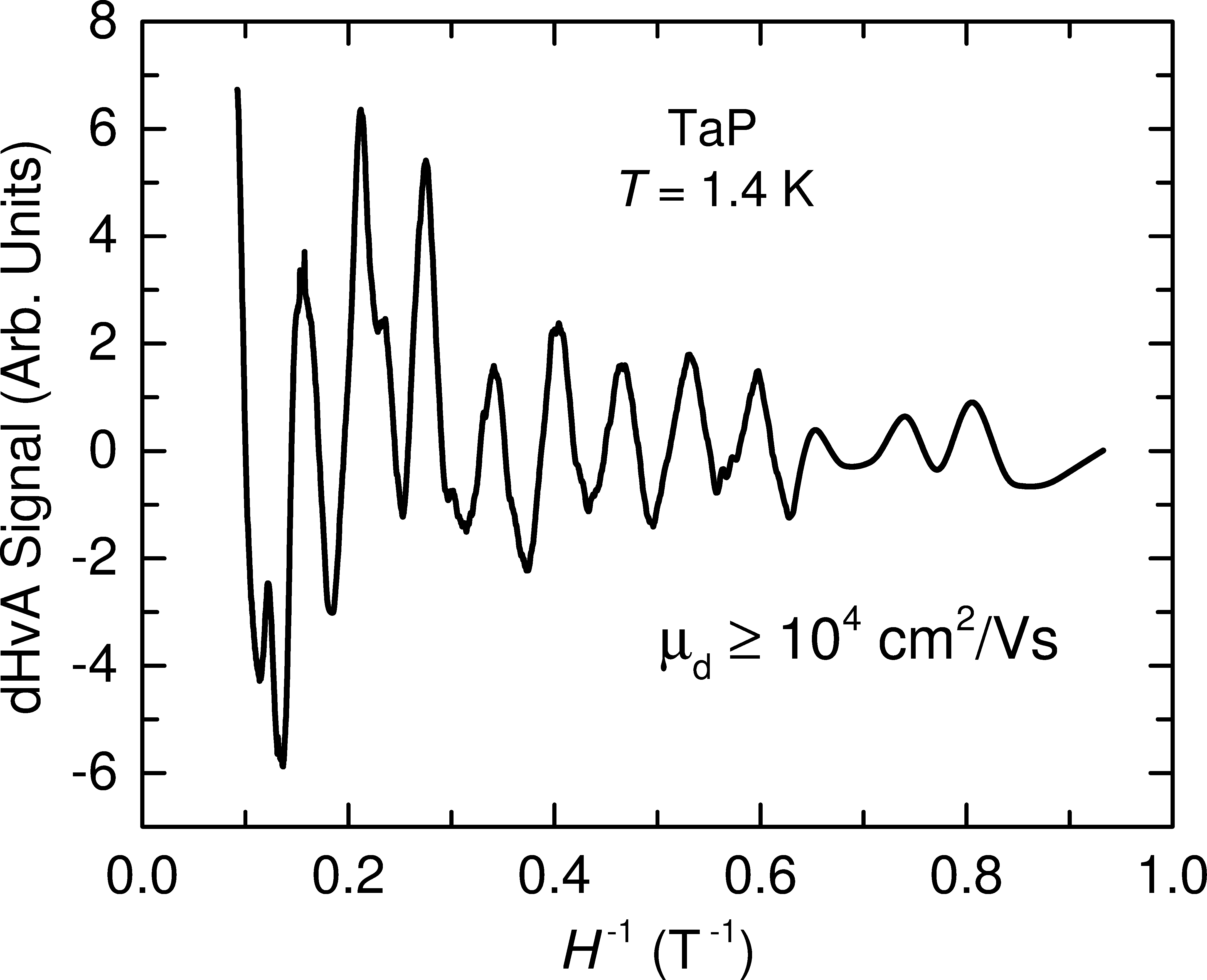}
    \end{center}
        \caption{The de Haas-van Alphen signal (via magnetic torque measurements) of TaP. The signal has been normalized by the field and the background has been subtracted to reveal only the oscillatory signal.}
        \label{fig:tapdhva}
\end{figure}

This non-stoichiometry initially hints at a large number of point defects (site vacancies) on the anionic P-sites, and could also be a sign of a potential superstructure, Ta$_6$P$_5$, in TaP, with 6 times the volume of the substructure. However, TEM/STEM images of a crushed crystal of TaP (Fig.~\ref{fig:taptem}) reveal a somewhat different picture: defects in the form of planar defects (stacking faults), anti-sites and vacancies. The planar defects are prevalent with a high density of stacking faults while point defects (anti-sites and vacancies) have much less density. In Fig.~\ref{fig:taptem}(a), the arrows indicate stacking faults along the $c$-axis. Fig.~\ref{fig:taptem}(b) is a STEM high angle annular dark field (HAADF) $Z$-contrast image which shows the atomic arrangement of Ta ($Z=73$, bright spheres) and P ($Z=15$, small low-intensity spheres). The image clearly displays the expected stacking of the known $I4_{1}md$ structure, disrupted by "shifts" (indicated by yellow lines) of one half lattice width in $a$ (or $b$), creating a region with a different stacking arrangement. Fig.~\ref{fig:taptem}(c) is an HAADF-STEM image of a different region clearly displaying anti-site regions. These are areas where the P sites are occupied by Ta and Ta sites occupied by P. The normally low-intensity P sites have a much higher intensity than expected in these regions, thus indicating a substantial number of Ta atoms in these locations. In addition, the Ta sites in these regions have a lower intensity as expected, a sign that there are P atoms in those columns. Furthermore, the anti-sites in Fig.~\ref{fig:taptem}(c) also show point defects in the form of Ta vacancies, an "empty" column of Ta atoms. It is possible that the column may contain some P atoms.

Fig.~\ref{fig:taptem}(d) shows an electron diffraction pattern of the single crystal in the $(h0l)$ plane with streaks along the $c^*$ direction for $h=2n+1$. Due to the stacking faults, the streaks appear along selected directions in reciprocal space and are pronounced in the $(h0l)$ plane with $h=2n+1$ positions (or, equivalently, in the $(0kl)$ plane with $k=2n+1$) only. The reason for this can be understood by examining the structure (Fig.~\ref{fig:struct}) and its layering of Ta atoms along the $c$-axis: as the structure is incrementally built up along $c$, Ta atoms are located in $(0,0,0)$, $(0,\frac{1}{2},\frac{1}{4})$, $(\frac{1}{2},\frac{1}{2},\frac{1}{2})$, $(\frac{1}{2},0,\frac{3}{4})$, and $(0,0,1)$, and so on for the next unit cell. One immediately notices that the vectors between these positions includes alternating shifts of $\frac{1}{2}a$ and $\frac{1}{2}b$ as the layers progress. When a stacking fault occurs, this alternating shift pattern is disrupted, resulting in the next Ta-atom to be stacked directly above the previous one. Additional layers of the same alignment may add on to create a slab consisting of a different stacking arrangement than the original stacking. To return to the original stacking, an additional fault would be needed in order to bring the next Ta-atom into the alternating shifts of $\frac{1}{2}a$ and $\frac{1}{2}b$ stacking pattern. Therefore, since the stacking faults involve shifts by $\frac{1}{2}a$ (or equivalently, $\frac{1}{2}b$), only $(h0l)$ with $h$ odd (or $k$ odd in $(0kl)$) show streaks.\\
\\
The clear evidence of defects and non-stoichiometry in TaP raises the question of crystal quality. However, magnetic torque measurements on a single crystal of TaP clearly display an oscillatory de Haas-van Alphen signal (Fig.~\ref{fig:tapdhva}). Oscillations are observed as low as 1 T, thus indicating a minimum drift mobility of $10^4$ cm$^2$/Vs at 1.4 K, comparable to other reports of ultrahigh carrier mobility in TaP.\cite{Arnold_ARXIV_2015} A Dingle plot\cite{Zeng_} suggests that the drift mobility even exceeds $10^5$ cm$^2$/Vs, despite the large number of defects. In fact, high frequency oscillations are also observed, most likely arising due a shift of the Fermi level produced by the defects and/or non-stoichiometry of the samples. Further, the oscillations differ from sample to sample, a feature that can be attribute to different stoichiometries for different crystals.\cite{Zeng_} The observed quantum oscillations are a sign that, electronically, these crystals are of comparable quality to those reported elsewhere, and it is likely that all experiments performed so far on these materials used non-stoichiometric samples with defects.\\
\\
\textbf{TaAs.} For TaAs, initially the same assumption was made as in TaP, \emph{viz.} a pnictide deficiency. However, the single crystal refinement clearly showed that the As site is \emph{not} deficient. In fact, refining the As site occupancy factor (SOF), it increased above 1. Fixing the arsenic SOF at 1.0 while refining the tantalum SOF, the occupancy of the Ta site, as expected, dropped to 0.92, yielding a stoichiometry of Ta$_{0.92}$As. In contrast, EDS analysis on different single crystals showed a wide range of elemental composition: from an As-deficiency (TaAs$_{0.91}$) to a Ta-deficiency (Ta$_{0.7}$As). This wide range observed with EDS and the discrepancy between the XRD and EDS measurements are the result of a high density of stacking faults, producing anti-domains.
\begin{figure*}[t]
    \begin{center}
        \includegraphics[width=1.0\textwidth]{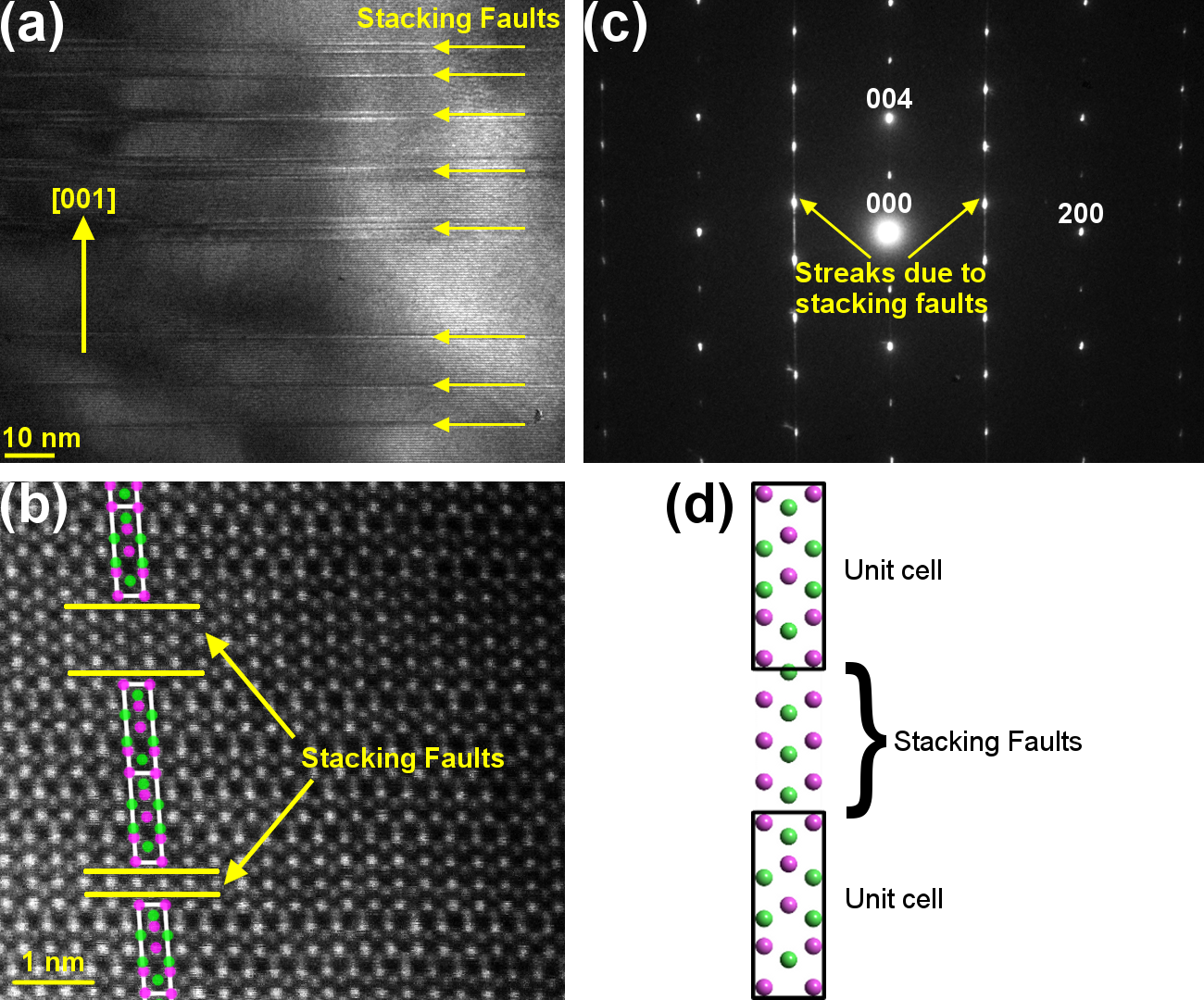}
    \end{center}
        \caption{(Color online) TEM/STEM images of TaAs (the tilt in the images is due to drift). (a) TEM bright field image view of a TaAs crystal. The faults in this case are also stacked along the $c$-axis. (b) Atomic resolution HAADF-STEM $Z$-contrast image viewed along $[010]$. A strip has been highlighted to indicate unit cells of the known $I4_{1}md$ structure and two regions of stacking faults. (c) Diffraction pattern of the $[010]$ direction with streaks arising due to the stacking faults. (d) Illustration of two unit cells separated by a region of stacking faults.}
        \label{fig:taastem}
\end{figure*}

TEM/STEM images of a crushed crystal of TaAs (Fig.~\ref{fig:taastem}) clearly show the high density of stacking faults. As in the case of TaP, the stacking faults occur along the $c$-axis (Fig.~\ref{fig:taastem}(a)). Fig.~\ref{fig:taastem}(b) shows the HAADF-STEM image where the Ta atoms are the bigger and brighter spheres. The atomic number of As ($Z=33$) is sufficient to show the As atoms as less bright and smaller spheres. An area along the $c$-axis has been highlighted to show where the structure is broken up by \emph{regions} of stacking faults (between the yellow lines). Fig.~\ref{fig:taastem}(c) shows the electron diffraction pattern in the $(h0l)$ plane of a TaAs single crystal, similarly to the TaP crystal, with streaks along the $c^*$ direction for $h=2n+1$. Fig.~\ref{fig:taastem}(d) is an illustration of the atomic arrangement showing two unit cells separated by stacking faults.\\
\\
Electronic structure calculations on stoichiometric TaAs, TaAs containing vacancies, and TaAs containing stacking faults were performed in order to elucidate the effects these structural features have on the Fermi surface, the Weyl points and the density of states (DOS).

The calculations confirm that Ta-deficient TaAs behaves as an electron-doped system and As-deficient TaAs as a hole-doped system when compared to stoichiometric TaAs (the band structures are displayed in the Appendix in Fig.~\ref{fig:TaAs_Vac_Bands}). Notice that the electronic structure is significantly altered when introducing vacancies: the position of the Weyl points relative to the Fermi level has shifted dramatically in the presence of vacancies when compared to the stoichiometric case. The DOS results are displayed in Fig.~\ref{fig:TaAs_DOS}. In the case of As vacancies, the DOS from $-2$~eV to 0~eV is still of similar shape to the DOS of stoichiometric TaAs near the $E_F$, but with non-zero states at $E_F$, as expected. In contrast, Ta vacancies distort the DOS at $E_F$ by about 0.5~states/eV/atom.

\begin{figure*}[t]
    \begin{center}
        \includegraphics[width=1.0\textwidth]{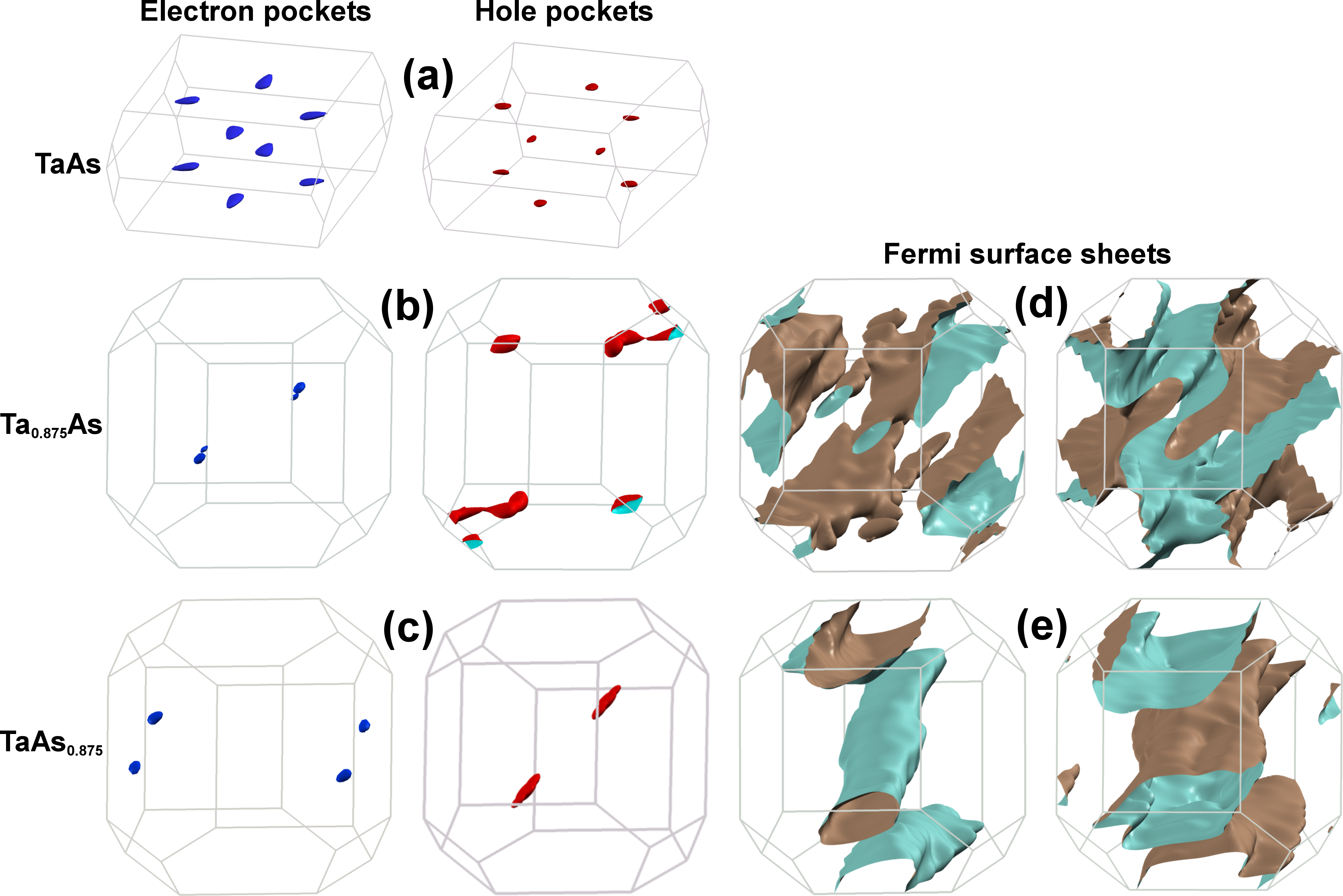}
    \end{center}
        \caption{(Color online) Fermi surfaces of TaAs, TaAs with a 12.5~\% Ta vacancy concentration, and TaAs with a 12.5~\% As vacancy concentration, calculated using a $2\times2\times1$ supercell (primitive unit cell). The electron and hole pockets are shown in blue and red color, respectively. In general the colors used in this figure are the same as used in Fig.~\ref{fig:TaAs_Vac_Bands} for the corresponding bands. Figure (a) shows the small electron and hole pockets in the case of TaAs without any vacancy, (b) the small electron and hole pockets that form when TaAs contains 12.5~\% Ta vacancies, and (c) the small electron and hole pockets that form when TaAs contains 12.5~\% As vacancies. The corresponding Fermi surface sheets obtained in the vacancy cases are shown in (d) and (e).}
        \label{fig:TaAs_Vac_FS}
\end{figure*}

Fig.~\ref{fig:TaAs_Vac_FS} shows the Fermi surfaces of three cases: stoichiometric TaAs, TaAs with 12.5~\% Ta vacancies, and TaAs with 12.5~\% As vacancies, calculated using a $2\times2\times1$ supercell (primitive unit cell), i.e., a supercell twice as long along the $a$- and the $b$-axis. A number of small electron and hole pockets of similar size to those found in the regular unit cell of stoichiometric TaAs exist (Fig.~\ref{fig:TaAs_Vac_FS}(a)), pockets associated with the Weyl points.\cite{Weng_PRX_2015, Huang_NatComm_2015} However, when introducing Ta and As vacancies in the form of point defects, the small electron and hole pockets appear to form \emph{away} from the Weyl points (Fig.~\ref{fig:TaAs_Vac_FS}(b) and (c)). Notice that in addition to the small pockets, much larger Fermi surface sheets emerge upon introducing vacancies (Fig.~\ref{fig:TaAs_Vac_FS}(d) and (e)). The presence of small electron/hole pockets can explain the similar behavior in Shubnikov-de Haas (SdH) oscillations where the larger frequencies may be harder to detect.\cite{Huang_PRX_2015}

\begin{figure}[t]
    \begin{center}
        \includegraphics[width=1.0\columnwidth]{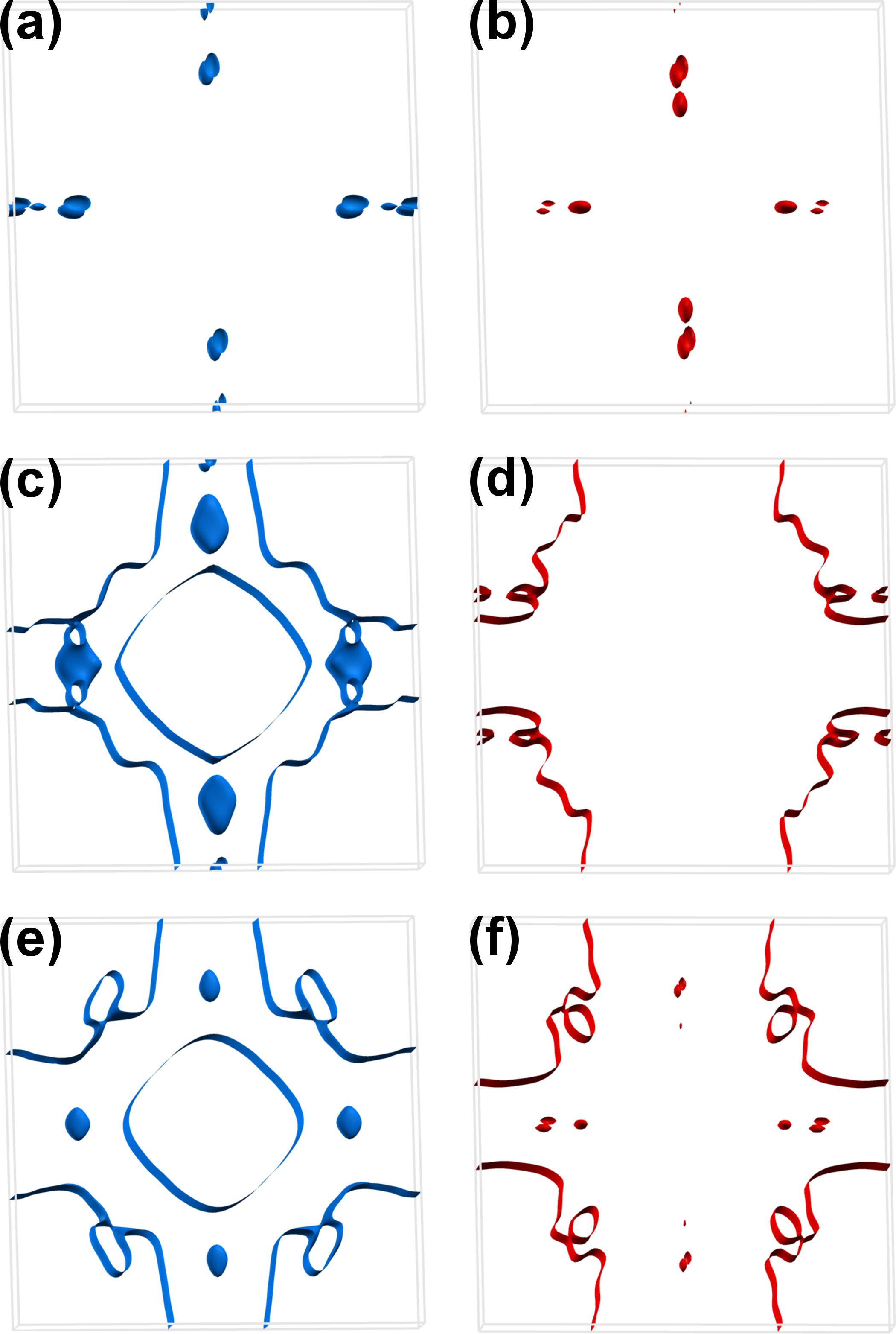}
    \end{center}
        \caption{(Color online) Figures (a) and (b) show the electron (blue) and hole (red) pockets which include the Weyl points in the case of a stoichiometric TaAs supercell which is three times the unit cell along the $c$-axis ($1\times1\times3$ supercell). Figures (c) and (d) show the corresponding Fermi sheets in the case of a TaAs $1\times1\times3$ supercell containing stacking faults. Figures (e) and (f) show the same electron and hole pockets as in (c) and (d) but with the Fermi level shifted by 60 meV upwards and downwards for the electron and hole pocket, respectively.}
        \label{fig:TaAs_SF_FS}
\end{figure}

However, as noticed from the TEM results (Fig.~\ref{fig:taastem}), the majority of defects are stacking faults, i.e., planar defects. Fig.~\ref{fig:TaAs_SF_FS} compares the Fermi surface sheets of stoichiometric, ideal, TaAs with that of a stoichiometric TaAs containing stacking faults (as depicted in Fig.~\ref{fig:taastem}(d)). The calculations were performed on a $1\times1\times3$ supercell, using a simple tetragonal unit cell (conventional unit cell) for the stoichiometric TaAs, and the stacking fault is accommodated by taking three times the size of this unit cell along the $c$-axis. If the conventional body-centered tetragonal unit cell (primitive unit cell) would be used, the smallest unit cell which includes a stacking fault would be larger. Figs.~\ref{fig:TaAs_SF_FS}(a) and (b) depicts the electron and hole pockets of the stoichiometric, ideal, supercell, which include the Weyl points. Figs.~\ref{fig:TaAs_SF_FS}(c) and (d) show the corresponding Fermi sheets when stacking faults are introduced. However, when shifting the Fermi level by a rather small amount of energy (60 meV upwards for the electron pockets and 60 meV downwards for the hole pockets) the pockets containing the Weyl points emerge again. Thus, while the calculation corresponds to a periodic array of stacking faults with a quite large stacking fault density of $1/3$, the Fermi surface structure is more or less preserved. Therefore, a different explanation for the robust nature of the SdH behavior over several samples may be that the main defects present in the crystals are stacking faults as opposed to As or Ta vacancies. Band structure calculations comparing stoichiometric, ideal, TaAs with TaAs containing stacking faults are displayed in Fig.~\ref{fig:TaAs_SF_Bands}. The effect of stacking faults near the Weyl point is to slightly move the Fermi level, but the main features of the band structure near the Weyl point are preserved.

Vacancy formation energy calculations were performed to shed some light on the defects in TaAs. Using a $2\times2\times1$ supercell (conventional body-centered unit cell), calculations for the compositions Ta$_{16}$As$_{15}$ (As vacancy) and Ta$_{15}$As$_{16}$ (Ta vacancy) were performed. The formation energy of an As vacancy was 4.09~eV per Ta$_{16}$As$_{15}$ unit, while only 2.90~eV per Ta$_{15}$As$_{16}$ unit for a Ta vacancy. This shows that the formation of Ta vacancies is more energetically favorable than the formation of As vacancies, in agreement with the results from the single crystal x-ray diffraction. In addition, a calculation of stacking fault energy yields $\Delta E=(E_{SF}-E_{0})/A=0.791$ eV/\textrm{\AA}$^{2}=126.7$ mJ/m$^{2}$. Here, $E_{SF}$ denotes the energy of a stacking fault structure, $E_{0}$ the energy of a perfect structure, and $A$ the surface area. The moderate value of the stacking fault energy suggest the likelihood of formation of stacking faults in these compounds.

\section{Conclusion}
We have shown that the two tantalum pnictide semimetals TaAs and TaP grow with stoichiometry deviations, and display a high number of defect densities in the form of stacking faults, anti-site disorder, and vacancies. The differences between the two samples is striking: while TaP displays the full range of defects and grows with a large anionic pnictide deficiency, TaP$_{0.83(3)}$, TaAs shows only stacking faults, although at a higher density than TaP, accompanied by cationic transition metal deficiency, Ta$_{0.92(2)}$As. As the two semimetals were grown using the same techniques reported in the literature, this indicates that, most likely, all experiments performed so far on these materials have used non-stoichiometric samples containing defects.

It is, however, clear that our crystals are electronically of comparable quality to those studied previously, despite the high number of defects and large stoichiometry variations, since our magnetic torque measurements on TaP reveal oscillatory de Haas-van Alphen signals and an ultrahigh carrier mobility, two observations normally associated with high quality crystals.

Our electronic structure calculations of TaAs show that while the Fermi surface is considerably altered when introducing vacancies, there still exist electron and hole pockets of similar size to those found in stoichiometric TaAs. These small electron and hole pockets, however, appear to form away from the Weyl points and the position of the Weyl points relative to the Fermi level is shifted. Introducing a periodic array of stacking faults show that the Fermi surface structure for the electron and hole pockets near the Weyl points is similar to the stoichiometric, ideal, TaAs, and the main effect of the stacking faults is to change the location of the Fermi level: by slightly raising or lowering the energy, Weyl fermion pockets appear again, and of the same size as seen in quantum oscillations. In other words, while vacancies significantly alter the location of the Weyl pockets, stacking faults seem to preserve the main features (such as location and dispersion) of the electronic structure near the Weyl points.\\
\\
With quantum oscillations experimentally observed in a single crystal containing a high number of stacking faults and defects, and Weyl fermion pockets appearing in the Fermi surface sheets of TaAs with stacking faults, it is clear that our results illustrate the robustness of these quantum phenomena in this family of semimetals.

Future studies of these materials need therefore to carefully analyze the stoichiometry, since it is clear that structural defects are present and may shift the actual stoichiometry. Interestingly, these defects do not seem to couple to the electronic transport properties, as reflected in the observed high mobilities, but do affect the Weyl point positions and should have consequences for the Weyl-related physics. It remains to be determined what these effects will be.

\section{Acknowledgments}
T.B. and T.S. are supported by the U.S. Department of Energy, Office of Basic Energy Sciences, Materials Sciences and Engineering Division, under Award \#DE-SC0008832. K.-W.C. and R.E.B. acknowledge support from the National High Magnetic Field Laboratory UCGP program. L.B. is supported by DOE-BES through Award \#DE-SC0002613. Work at the University of Missouri (J.S. and D.J.S) is supported by the U.S. Department of
Energy, Basic Energy Sciences through the S3TEC Energy Frontier Research Center, Award No.
DE-SC0001299/DE-FG02-09ER46577. This work was performed at the National High Magnetic Field Laboratory, which is supported by the NSF cooperative agreement DMR-1157490 and the State of Florida.

\appendix
\section{Band Structures and Density-of-States}
\subsection{Vacancies}
\begin{figure*}[t]
    \begin{center}
        \includegraphics[width=1.0\textwidth]{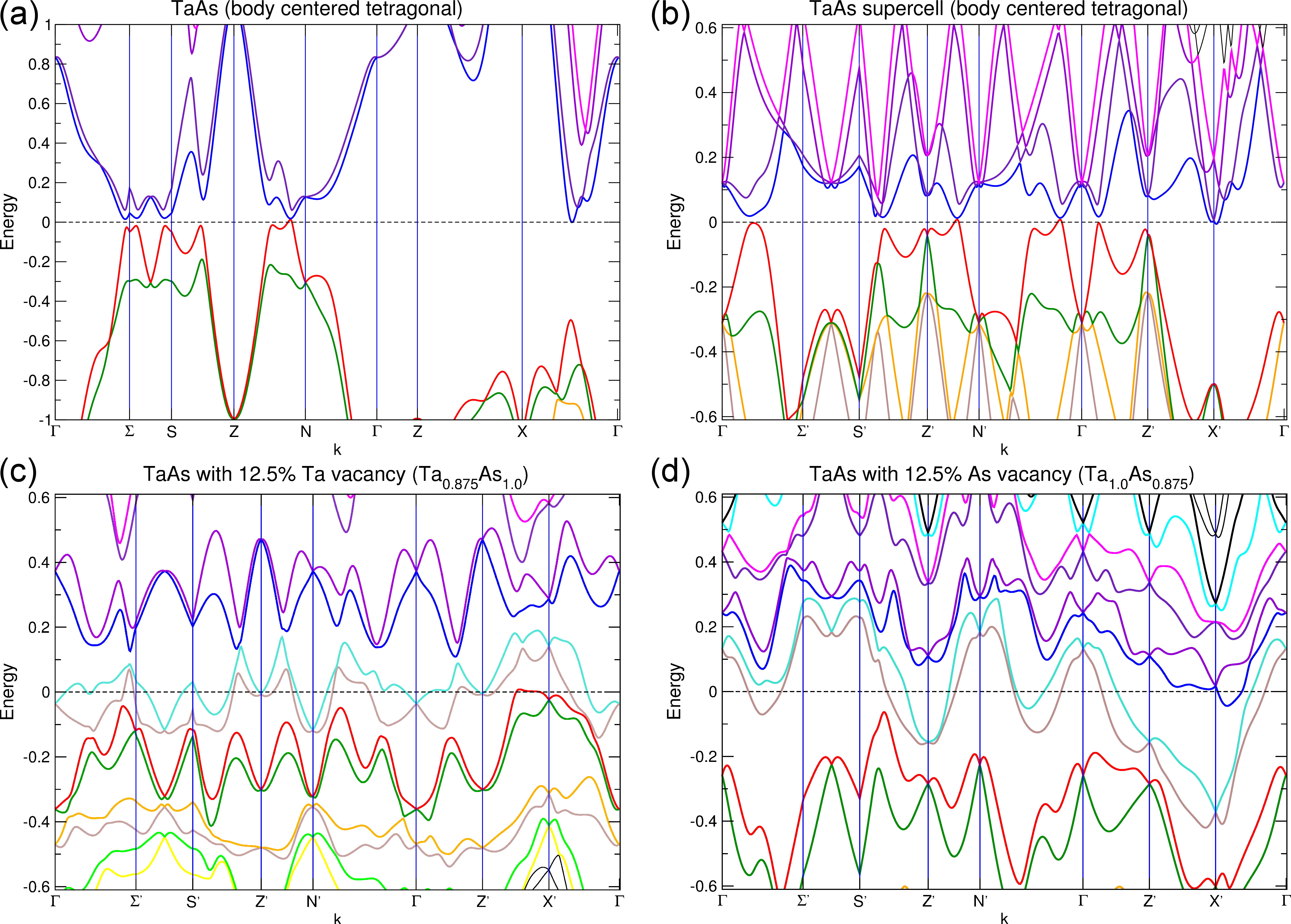}
    \end{center}
        \caption{(a) The band structure of TaAs in the body centered tetragonal phase consistent with published results. (b) The band structure of TaAs supercell containing four times the volume of the unit cell, twice along the $a$- and the $b$-axis. (c) The band structure of TaAs with  12.5~\% Ta vacancy concentration. (d) The band structure of TaAs with 12.5~\% As vacancy concentration.}
        \label{fig:TaAs_Vac_Bands}
\end{figure*}

Figure~\ref{fig:TaAs_Vac_Bands} displays the results of band structure calculations of TaAs in the case of introduced vacancies. Fig.~\ref{fig:TaAs_Vac_Bands}(a) shows the band structure of TaAs obtained in the body-centered tetragonal phase which is consistent with published results. Fig.~\ref{fig:TaAs_Vac_Bands}(b) reproduces the band structure of the stoichiometric TaAs using a supercell twice as long along the $a$- and the $b$-axis (four times the volume of the unit cell), i.e., a $2\times2\times1$ supercell (primitive unit cell). This is in order to identify the pockets near the Weyl points and to compare the band structure of the stoichiometric case with the cases of 12.5~\% vacancy concentration which require a similar supercell. Fig.~\ref{fig:TaAs_Vac_Bands}(c) shows the results obtained for the band structure of TaAs with 12.5~\% Ta vacancy concentration, while Fig.~\ref{fig:TaAs_Vac_Bands}(d) shows the band structure of TaAs with 12.5~\% As vacancy concentration. The band structures are consistent with the fact that TaAs with Ta vacancies behaves as an electron-doped system whereas TaAs with As vacancies behaves as a hole-doped system.\\
\\

\begin{figure}[htb]
    \begin{center}
        \includegraphics[width=1.0\columnwidth]{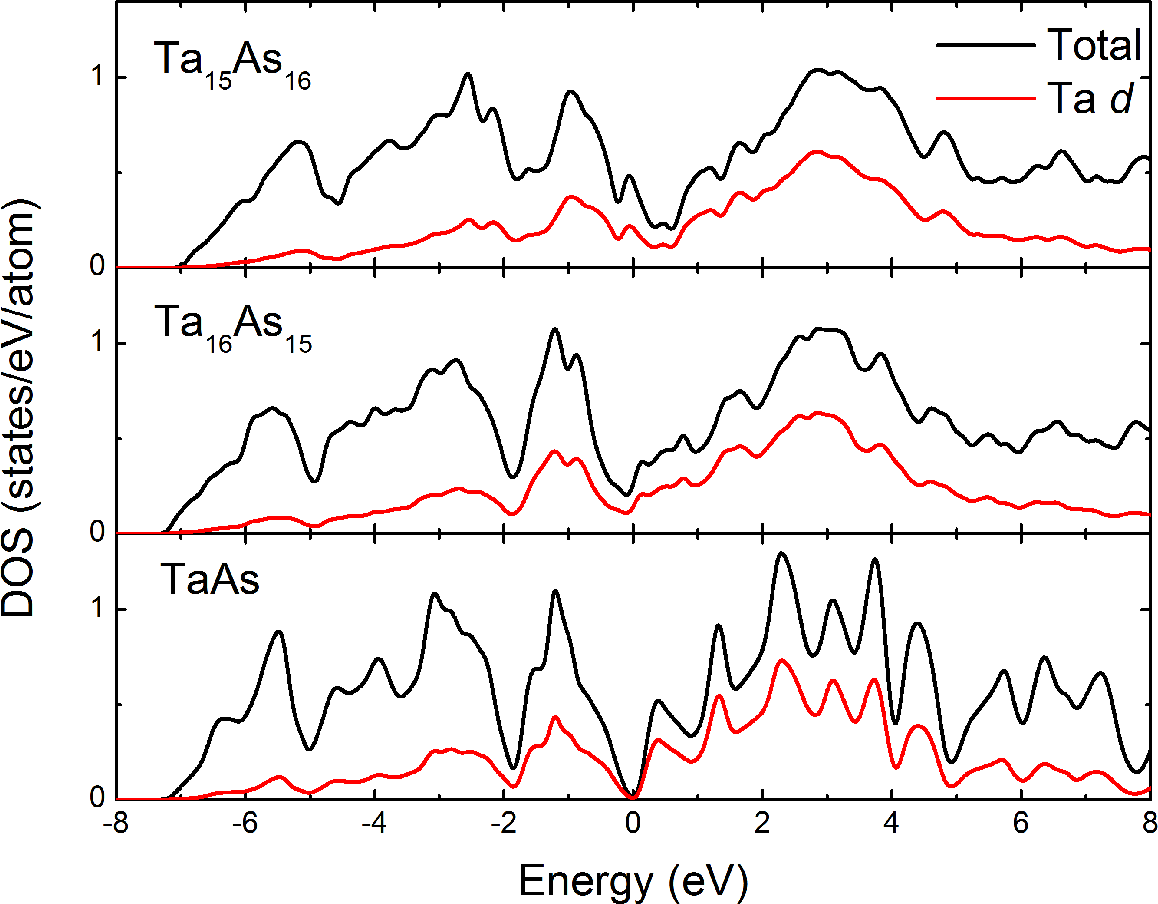}
    \end{center}
        \caption{Total and partial density of states of stoichiometric TaAs (bottom panel), TaAs with As vacancies (middle panel), and TaAs with Ta vacancies (top panel).}
        \label{fig:TaAs_DOS}
\end{figure}

Figure~\ref{fig:TaAs_DOS} shows the total and partial density of states (DOS) of stoichiometric TaAs, along with slightly Ta-vacant TaAs, Ta$_{15}$As$_{16}$, and As-vacant TaAs, Ta$_{16}$As$_{15}$, calculated using a $2\times2\times1$ supercell (conventional unit cell). The As vacancy case has a DOS that is still of similar shape to the DOS of stoichiometric TaAs near the $E_F$, from $-2$~eV to 0~eV but with non-zero states at $E_F$, as expected. In contrast, the Ta vacancy case seems to have distorted the DOS near $E_F$, with about 0.5~states/eV/atom at $E_F$.

\subsection{Stacking Faults}
\begin{figure*}[htb]
    \begin{center}
        \includegraphics[width=1.0\textwidth]{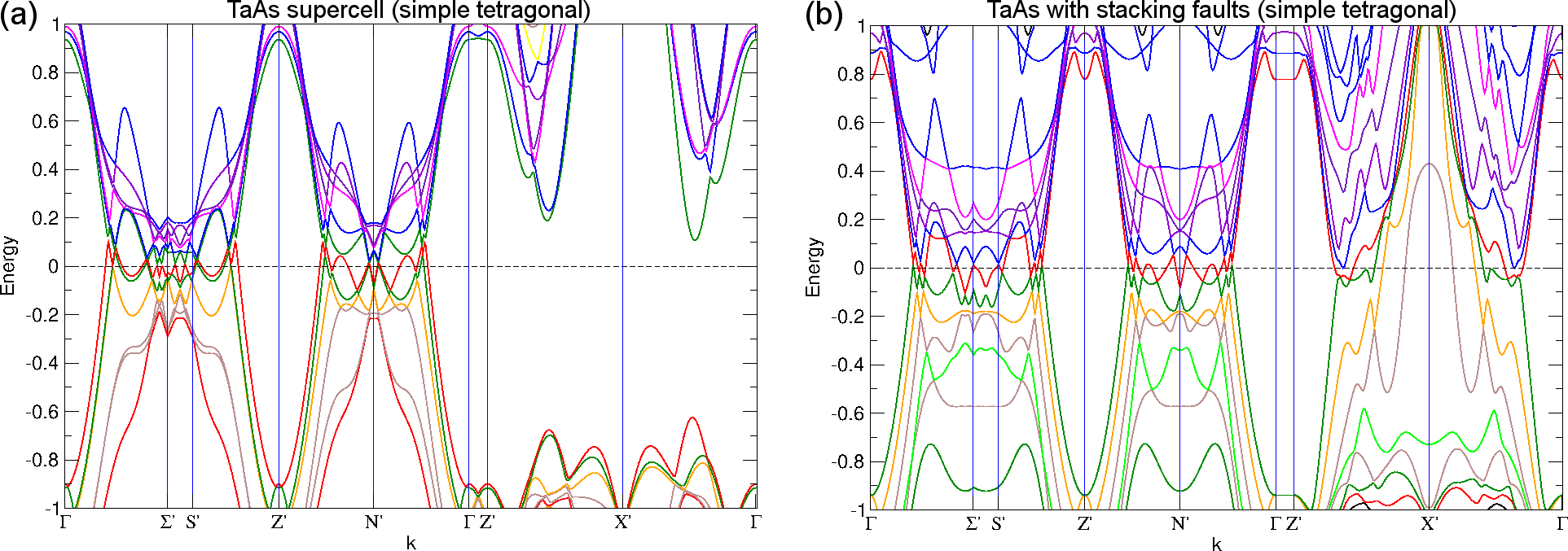}
    \end{center}
        \caption{(a) The band structure along certain high symmetry points of a stoichiometric TaAs supercell three times the unit cell along the $c$-axis. (b) The band structure in the case of a TaAs supercell containing one stacking fault.}
        \label{fig:TaAs_SF_Bands}
\end{figure*}

Figure~\ref{fig:TaAs_SF_Bands} compares the band structure of a stoichiometric, ideal, TaAs to that of TaAs containing a stacking fault. In these calculations, a simple tetragonal unit cell for the stoichiometric TaAs was used and the stacking fault are accommodated by taking three times the size of this unit cell along the $c$-axis, a $1\times1\times3$ supercell (conventional unit cell). The main effect of the stacking fault is the significant changes in the bands in directions $\Gamma\to X^{\prime}$ and $X^{\prime}\to Z^{\prime}$ which do not contain the Weyl point (i.e., $\Sigma^{\prime}$). The effect of the stacking fault near the Weyl point, however, is mainly to slightly move the Fermi level, and the main features of the band structure near the Weyl point are preserved.

\section{Methods}
\textbf{Sample Preparation.} Single crystals of TaAs were grown by chemical vapor transport, as previously described.\cite{Ghimire_JPCM_2015} Polycrystalline precursor specimens were first prepared by sealing elemental Ta and As mixtures under vacuum in quartz ampoules and heating the mixtures at a rate of 100$\degree$C/hr to 700$\degree$C, followed by a dwell at this temperature for 3 days. The polycrystalline TaAs boules were subsequently sealed under vacuum in quartz ampoules with 3 mg/cm$^3$ of iodine to serve as the transporting agent. The ampoules had diameters 1.4 cm and lengths 10 cm and were placed in a horizontal tube furnace such that a temperature gradient would be established during firing. The ampoules were slowly heated at a rate of 18$\degree$C/hr, and put into a temperature gradient with $\Delta T=850\degree$C$-950\degree$C. The ampoules were maintained under this condition for 3 weeks and were finally rapidly cooled to room temperature. This process produced a large number of single crystal specimens with typical dimensions of 0.5 mm on a side.

TaP single crystals were synthesized through a chemical vapor transport technique using iodine as the transport agent. 99.98\% pure Ta powder and 99\% pure P lumps were introduced into quartz tubes together with 99.999\% pure iodine serving as the transporting agent. The quartz tubes were evacuated, sealed, brought to 500$\degree$C, held at this temperature for 1 day, then brought to 650$\degree$C, held for 12 hours, and then finally raised to 975$\degree$C and held there for 5 days. Subsequently, they were cooled to 800$\degree$C and held there for 1 day, followed by an air quench.\\
\\
\textbf{Single Crystal X-ray Diffraction.} Crystals of the semimetals were structurally characterized by single crystal x-ray diffraction using an Oxford-Diffraction Xcalibur2 CCD system with graphite-monochromated Mo$K\alpha$ radiation. Data was collected to a resolution of 0.4~\textrm{\AA}, equivalent to $2\theta = 125\degree$. Reflections were recorded, indexed and corrected for absorption using the Agilent CrysAlisPro software.\cite{CrysAlisPro} Subsequent structure refinements were carried out using CRYSTALS,\cite{Crystals} using atomic positions from the literature.\cite{Pearson} The data quality for all samples allowed for an unconstrained full matrix refinement against $F^2$, with anisotropic displacement parameters for all atoms. Crystallographic information files (CIFs) have been deposited with ICSD (CSD Nos. 430436 and 430437 for TaP and TaAs, respectively).\cite{ICSD}\\
\\
\textbf{EDS.} EDS was performed with a field-emission scanning electron microscopy (Zeiss 1540 XB), on 6 to 12 spots each on the several single crystals studied. The EDS stoichiometries quoted here result from average values.\\
\\
\textbf{TEM.} The TEM samples were prepared by crushing single crystals that were previously checked by XRD, in Ethyl Alcohol 200 Proof in a pestle and mortar. The suspension was then dropped onto a carbon/formvar TEM grid (Ted Pella, Inc.) using a 1.5 ml pipette. TEM/STEM images were collected using the probe aberration corrected JEOL JEM-ARM200cF with a cold field emission gun at 80 kV to avoid beam damage. The STEM high angle annular dark field (STEM-HAADF) images were taken with the JEOL HAADF detector using the following experimental conditions: probe size 7c, CL aperture 30 $\mu$m, scan speed 32 $\mu$s/pixel, and camera length 8 cm, which corresponds to a probe convergence semi-angle of 11 mrad and collection angles of $76-174.6$ mrad. Qualitatively, the intensity of atomic columns in STEM-HAADF images is proportional to the atomic number $Z^n$, where $n$ is close to 2, i.e., they are $Z$-number-sensitive images ($Z$-contrast). The STEM resolution of the microscope is 0.78~\textrm{\AA}.\\
\\
\textbf{Electronic structure calculations.} Electronic structure calculations were performed by using the Vienna \emph{ab-initio} simulation package\cite{Shiskin_PRL_2007, Fuchs_PRB_2007, Shiskin_PRB_2007, Shiskin_PRB_2006} (VASP) within the generalized gradient approximation (GGA). We have included the contribution of spin-orbit coupling in our calculations. The Perdew-Burke-Ernzerhof (PBE) exchange correlation functional \cite{Perdew_PRL_1996} and the projected augmented wave (PAW) methodology\cite{Blochl_PRB_1994} were used to describe the core electrons. The 5\emph{s}, 5\emph{p}  and 5\emph{d} electrons for Ta and the 3\emph{d}, 4\emph{s} and 4\emph{p} electrons for As were treated as valence electrons in all our calculations. The energy cut off for the plane-wave basis was chosen to be 600 meV. A total of 208 bands and a $k-$point mesh of $8\times8\times8$ were used for the self-consistent ground state calculations. A total of 100 $k-$points were chosen between each pair of special $k-$points in the Brillouin-zone for the band-structure calculations. The Fermi surfaces were generated using a $k-$point mesh of $10\times10\times16$. The Fermi surfaces were generated using the eigenvalues obtained from VASP and were visualized using the XCrysden software.\cite{Kokalj_CMS_2003}

In the case of the formation energy calculations, the calculations were carried out using the linearized augmented plane-wave (LAPW) method
as implemented in the WIEN2K code, with the fully relaxed structures generated from VASP. The LAPW sphere radii were 2.35 bohr for both Ta and As. The cut-off parameter for the basis was $R_{min}K_{max}=9$. We used well converged $k$-point sampling for the total energy calculations which
are sensitive to $k$-points, especially in the formation energy calculations. For the vacancy formation energy calculation, a $2\times2\times1$ supercell was generated, which includes 31 Ta and As atoms and one Ta or As vacancy, yielding a composition of Ta$_{15}$As$_{16}$ and Ta$_{16}$As$_{15}$, respectively. Then the vacancy formation energy was calculated using $\Delta H_{f}=E_{\alpha}-E_{\textrm{host}}+\sum_{\alpha}n_{\alpha}\mu_{\alpha}$, where $E_{\alpha}$ and $E_{\textrm{host}}$ are the energies with and without vacancy $\alpha$. Here, $n_{\alpha}$ and $\mu_{\alpha}$ are the number of vacancies and the chemical potential of vacancy $\alpha$ in the elemental phase, respectively. The formation energies shown in the results are calculated within WIEN2K with spin-orbit coupling.

\newpage
\bibliography{TaPn_bib}
\end{document}